\documentclass[12pt]{article}%
\usepackage{amsfonts,amsmath}
\usepackage{amssymb}
\usepackage{dsfont}
\usepackage{mathtools}

\makeatletter \@addtoreset{equation}{section} \makeatother

\newcommand{\WMO}{M}

\newcommand{\tr}{\mathbb{D}}%   {\tr }

\newcommand{\wmv}{\Omega}
\newcommand{\Go}{\Omega}
\newcommand{\bwmv}{{\bar{\Omega}}}
\def\Ex{{\E(\Omega)}}

\newcommand{\dga}{{\dot{\ga}}}
\newcommand{\ff}{\frac }
\newcommand{\gT}{\mathcal{T}}
\newcommand{\gt}{\tau}
\newcommand{\bgt}{\bar{\tau}}
 \newcommand{\eq}{\eqref}

\newcommand{\hhmt}{{{h}}}

\newcommand{\hmt}{{\vartriangle}}

 \newcommand{\be}{\begin{equation}}
\newcommand{\ee}{\end{equation}}
\newcommand{\bee}{\begin{eqnarray}}
\newcommand{\beee}{\begin{array}}
\newcommand{\eee}{\end{eqnarray}}
\newcommand{\eeee}{\end{array}}
\newcommand{\gn}{\nu}
\newcommand{\EE}{\mathcal{E}}

\newcommand{\bEE}{\bar{ \mathcal E }}

\newcommand{\gx}{\xi}
\newcommand{\gf}{\zeta}%{\phi}!!!!!!!!!!

\newcommand{\ga}{\alpha}
\newcommand{\pa}{{\dot{\ga}}}
\newcommand{\pb}{{\dot{\gb}}}

\newcommand{\pga}{{\bar{\gamma}}}
\newcommand{\dgga}{{\dot{\gamma}}}

\newcommand{\gb}{\beta}
\newcommand{\gga}{\gamma}

\newcommand{\M}{{\cal M}}

\newcommand{\E}{{\cal E}}

\newcommand{\W}{{\cal W}}

\newcommand{\rhs}{{\it r.h.s.} }

\newcommand{\lhs}{{\it l.h.s.} }

\newcommand{\ie}{{\it i.e.,} }
\newcommand{\ls}{\!\!\!\!\!\!}

\newcommand{\gd}{\delta}

\newcommand{\gk}{\kappa}
\newcommand{\gep}{\epsilon}
\newcommand{\gvep}{\varepsilon}

\newcommand{\gs}{\sigma}

\newcommand{\bz}{\bar z}
\newcommand{\go}{\omega}

\newcommand{\by}{{\bar{y}}}
\newcommand{\brr}{{\bar{r}}}

\newcommand{\bk}{{\bar{k}}}

\newcommand{\bp}{{\bar{p}}}

\newcommand{\q}{\,,\qquad}

\newcommand{\dgb}{{\dot{\beta}}}

\newcommand{\nn}{{\nonumber}}

\newcommand{\half}{\frac{1}{2}}

\newcommand{\p}{\partial}

\newcommand{\D}{{\cal{D}}}

\newcommand{\B}{{  B}}%%

 \newcommand{\dr}{{\rm d}}
\newcommand{\PP}{ \mathcal{ P} }

\begin{document}
\begin{flushright}
FIAN/TD/2025-17\\
\end{flushright}

\vspace{0.5cm}
\begin{center}
{\large\bf  Lorentz Covariance of the $4d$ Nonlinear Higher-Spin Equations via BRST}

 \vglue 0.6  true cm
 %\vskip2cm HERE

\vskip0.5cm

 O.A. Gelfond$^1$ and M.A.~Vasiliev$^{1,2}$
 \vglue 0.3  true cm

\vspace{1 cm}

  \vspace{0.5cm}

 \textsc{$^{1}$} \textit{I.E. Tamm Department of Theoretical Physics,
Lebedev Physical Institute,}\\
 \textit{ Leninsky prospect 53, 119991, Moscow, Russia}\\

\par\end{center}

\begin{center}
\textsc{$^{2}$}\textit{Moscow Institute of Physics and Technology,}\\
 \textit{ Institutsky lane 9, 141700, Dolgoprudny, Moscow region,
Russia}\\

\par\end{center}

 \vspace{1.5cm}

{\it $\phantom{MMMMMMMMMMMMMMMMMM}$ To the memory of our friend and colleague \\
$\phantom{MMMMMMMMMMMMMMMMMMMMMMMMMMMMMM}$Simeon Konstein}

\vspace{1.5 cm}

\begin{abstract}
 \end{abstract}

We propose a BRST extension of the higher-spin gauge theory in $AdS_4$ with the
BRST operator associated with the  local Lorentz symmetry. Our construction supports
 manifest local Lorentz covariance  and
 is applicable both to any homotopy scheme of the perturbative analysis including the recently proposed differential homotopy and to the variety of
 further extended higher-spin models  in $AdS_3$ and $AdS_4$ with higher differential forms and Coxeter higher-spin algebras.

\newpage

  \tableofcontents

\newpage
\section{Introduction}
\label{Motivation}

The Cartan formulation of gravity in terms of differential forms is
based on the two fundamental symmetry principles: diffeomorphisms and local Lorentz symmetry. The
latter acts on all Lorentz tensor-spinors and reduces the formulation in terms of
the vielbein one-form to that in terms of the metric. To consider a theory of gravity
interacting with any matter it is crucially important to respect both of
these symmetries. The formulation in terms of exterior algebra automatically respects
diffeomorphisms. In particular, this is true in the unfolded formulation of the
nonlinear higher-spin (HS) gauge theory of \cite{Vasiliev:1992av}. The situation with local Lorentz symmetry is more tricky.
Though, naively, the formulation of \cite{Vasiliev:1992av} is $\mathfrak{sl}(2|\mathbb{C})$ invariant, this $\mathfrak{sl}(2|\mathbb{C})$ is
not manifest due to a non-zero
vacuum expectation value of one of the fields. That the full nonlinear equations of
\cite{Vasiliev:1992av} respect the local   Lorentz symmetry
 is a delicate property  originally shown in \cite{Vasiliev:1990vu}. As explained in \cite{Vasiliev:Rev} (see also \cite{Prokushkin:1998bq}),
after all it follows from the  special form of the HS equations related to
the deformed oscillator algebra of \cite{Vasiliev:1989qh,Vasiliev:1989re} providing an alternative
form of the Wigner oscillator \cite{wig}.

Though HS theory is proven to be Lorentz covariant, the problem of finding proper
variables in which it is manifestly covariant is involved and has been studied by
a number of authors within different formalisms \cite{different formalism1}-\cite{different formalism5}. An essential progress was made
in \cite{Didenko:2017qik} where an explicit construction was found for a particular formulation of
the theory resulting from the so-called conventional homotopy approach which, being the
simplest, is not necessarily appropriate from the locality perspective (see e.g.
\cite{Didenko:2018fgx}). In this paper we present a new approach allowing to build
manifestly Lorentz covariant formulation of the HS gauge theory in any homotopy scheme including the recently proposed
differential homotopy \cite{Vasiliev:2023yzx}. It is analogous to the
BRST
 extended version of the HS equations proposed recently
  in \cite{Vasiliev:2025hfh}  to formulate the $\mathfrak{sp}(2)$ invariance condition  and factorization transformations in the $d$-dimensional HS models.
  In this paper we use
the BRST language to Lorentz covariantize the spinor form of HS equations in $AdS_4$
with   the BRST sector directly associated with the manifest local Lorentz symmetry.
This leads to an additional shift symmetry acting both on
 the spin two Lorentz connection $\go^L$ of the original HS theory and on a new Lorentz
 connection  $\hat \go$. By construction, the new Lorentz connection $\hat\go$ acts
  in a manifestly Lorentz covariant way on all multispinors.  Therefore, the gauge condition
 $\go^L=0$ with respect to the shift symmetry leads to a manifestly Lorentz covariant formulation of the theory.

 The layout of the rest of the paper is as follows.
In Section \ref{HSsketch} we recall the form of the HS  equations  and the shifted  and  differential homotopy
approaches to their perturbative analysis.
 The  Ansatzes for the linear in $\eta$ ($\bar \eta$)     deformations  are also recalled. In Section \ref{LCHS} the derivation
of the Lorentz covariant HS  (LCHS) equations via BRST formalism is  presented,  the
gauge symmetries of the resulting system are considered and a part of them  is shown to
     form  a  Stueckelberg  symmetry allowing  to
       formulate   the theory in the manifestly Lorentz covariant form.
In   Sections \ref{VacQ} and \ref{Perturbative analysis} the vacuum solution and the linearized analysis of
LCHS equations are described within the shifted homotopy approach.
The differential homotopy approach is elaborated in Section \ref{LCDifHom} where it
is argued to be most convenient in the BRST formalism.
Conclusions   are in Section \ref{Con}.

\section{Original  higher-spin equations}
 \label{HSsketch}
\subsection{General setup}
HS equations in $AdS_4$
of \cite{Vasiliev:1992av} (see also \cite{Vasiliev:Rev}) have the   form
\begin{align}
&\dr\W+\W*\W=i\theta^{A}\theta_{A}+i\eta \B*\gga+i\bar{\eta}
\B*\bar{\gga}
\,,\label{hs1}\\
&\dr \B+[\W, \B]_*=0\,,\label{hs2}
\end{align}
where $\W(Z,Y; K |x)$ is a one-form in the double graded space
equipped with anticommuting differentials $\dr x^{\underline{m}}$,\, $\dr:=\dr x^{\underline{m}}\ff{\p}{\p x^{\underline{m}}}$ and
$\theta^A:=\dr Z^A$,
\begin{align}
&\W(Z,Y; K |x)=W_{\underline{m}}(Z,Y; K |x)\dr
x^{\underline{m}}+S_{A}(Z,Y; K |x)\theta^A\,,\\
&\{\dr x^{\underline{m}}, \dr x^{\underline{n}}\}=\{\dr
x^{\underline{m}}, \theta^A\}=\{\theta^A, \theta^B\}=0\,.
\end{align}
The field $\W$ contains both the HS potentials
(Fronsdal fields of spins $s\geq 1$) along with
their descendants in   $W_{\underline{m}}$ and the
compensator-like field $S$ in the $\theta$-subsector. A zero-form $\B$ contains
lower-spin matter fields and gauge covariant field strengths of the  HS  fields \cite{Vasiliev:1988sa}.
Space-time indices $\underline{m}, \underline{n}$ as well as
Majorana fiber spinor  indices $A, B$ take four values.

Apart from space-time coordinates $x^{\underline{m}}$, $\W$ and $\B$ depend on a number of
auxiliary variables. Commuting twistor-like variables
$Y_{A}=(y_{\ga}, \bar{y}_{\pa})$ and $Z_{A}=(z_{\ga}
,\bar{z}_{\pa})$ are designed to pack towers of HS fields and their descendants
into generating functions $\W$ and $\B$.
Spinor indices $\ga, \gb,...$ take two values as well as $\dga,\dgb$.

The associative star product $*$ acts on functions
$f(Z,Y)$
\be\label{star}
(f*g)(Z,Y)=\ff{1}{(2\pi)^4}\int dU dV f(Z+U, Y+U)g(Z-V,
Y+V)e^{iU_A V^A}\,,
\ee
where $U_AV^A:=U_A V_B \gep^{AB}$ with some $sp(4)$-invariant symplectic form
\be \gep_{AB}=-\gep_{BA}\q \gep_{AB}=(\gep_{\ga\gb},\gep_{\pa\pb})\,.\ee
 Indices are raised and lowered   as follows, $X^{A}=\gep^{AB}X_{B}$ and
$X_{A}=\gep_{BA}X^{B}$. The star product yields the elementary
commutation relations
\be\label{[Y*Y]}
[Y_A ,Y_B]_*=-[Z_{A}, Z_{B}]_*=2i\gep_{AB}\,,\qquad [Y_A,
Z_B]_*=0\,.
\ee
It admits the inner Klein operators
\be
\gk=e^{iz_{\ga}y^{\ga}}\,,\qquad \bar\gk=e^{i\bar z_{\pa}\bar
y^{\pa}}\,
\ee
with the characteristic properties
\be\label{Klpr}
\{\gk, y_{\ga}\}_*=\{\gk, z_{\ga}\}_*=0\,,\qquad \gk*\gk=1\,,
\ee
and analogously in the antiholomorphic sector for $\bar{\gk}$.

To distinguish between the adjoint representation for HS
potentials and the twisted-adjoint for HS curvatures as well as to
have a room for topological degrees of freedom in the HS system one
introduces Clifford-like outer Klein operators $K=(k, \bar k)$.
$k$ anticommutes with the holomorphic
variables $y_{\ga}, z_{\ga}, \theta_{\ga}$ and commutes with all
antiholomorphic ones, \bee\label{Klprout}
\{ k, y_{\ga}\}_*=\{ k, z_{\ga}\}_*=\{ k, \theta_{\ga}\}_*=0\q
[ k, \by_{\pa}]_*=[ k, \bz_{\pa}]=[ k, \bar\theta_{\pa}]_*=0\,.
\eee
 Analogously, $\bar k$ anticommutes with
the antiholomorphic variables and commutes with the holomorphic ones. In
addition,
\be
\label{kk}
k*k=\bar k *\bar k=1\,,\qquad [k, \bar k]_*=0.
\ee
The latter relations make the dependence on the Klein operators at
most bilinear
\be
\W=\sum_{i,j=0,1}\W_{i,j}k^i\bar k^{j}\,,\qquad
B=\sum_{i,j=0,1}B_{i,j}k^i\bar k^{j}\,.
\ee
Dynamical HS fields are contained in $\W(-k,-\bar k)=\W(k, \bar k)$ and
$B(-k,-\bar k)=-B(k, \bar k)$. The rest decompose into an infinite tower
of fields,  carrying at most a finite number of
degrees of freedom each, being therefore topological.

 Note that being analogous to inner  operator $\gk$, $k$   admits no star-product
realization because of its anticommutativity with
$\theta^{\ga}$.  More precisely, formulae (\ref{Klprout}), (\ref{kk}) extend
the star product (\ref{star}) to the outer Klein operators.

There are two types of central elements on the \rhs of
\eqref{hs1}. One is $\theta^ A\theta_A$ which obviously
commutes with any variable. The other ones are
\be
\label{gamma}
\gga=2k\gk\gd^{2}(\theta)\,,\qquad
\gd^2(\theta):=\ff12\theta^{\ga}\theta_{\ga}\q
\pga=2\bk\bar\gk\gd^{2}(\bar\theta)\,,\qquad
\gd^2(\bar\theta):=\ff12\bar\theta^{\pa}\bar\theta_{\pa}\,,\ee
which commute with everything including $\theta^{\ga}$, $\bar\theta^{\pa}$ thanks to
the Grassmann $\delta$-function, $\theta^{\ga}\gd^{2}(\theta)=0$.
 The complex constant
$\eta=|\eta| e^{i\phi}$ is the only free parameter of the $4d$ HS theory
with the \rhs of \eqref{hs1} linear in $B$.
For the parity preserving cases with $\phi=0$ and $\phi=\pi/2$ the theory is known as
the A-- and  B--model, respectively \cite{SS}.

In terms of $\theta^A $-independent and $\dr x$-independent  components  $W$ and $S$, respectively, of  $\W$ \eqref{hs2} reads as
\begin{align}
&\dr W+W*W=0\,,\label{HS1}\\
&\dr B+[W,B]_*=0\,,\label{HS2}\\
&\dr S+\{W,S\}_*=0\,,\label{HS3}
\\&S*S=-i\theta_{\ga}\wedge \theta^{\ga}(1+\eta B*k\gk)-
i\bar\theta_{\pa}\wedge \bar\theta^{\pa}(1+\bar\eta B*\bar k\bar \gk)\,,\label{HS4}\\
&[S,B]_*=0\,\label{HS5}
 . \end{align}

For the future convenience  let us
  define following   \cite{Vasiliev:1989qh,Vasiliev:1989re} the    dynamical
   $\mathfrak{sl}(2|\mathbb{C})-$generator
   $\WMO_{AB}$   via the field  $S=S_\ga\theta^\ga+S_\pa\bar \theta^\pa$.
    Introducing additional operators $\rho$   and   $\bar\rho$     \cite{Prokushkin:1998bq,Didenko:2017qik},
\bee
\rho k+k\rho=0\,, \qquad
\rho^2=1\,\q
\bar\rho \bar k+\bar k\bar\rho=0\,,\qquad  \bar\rho^2=1\,
,
\eee
that commute with the rest elements,   defining  derivations
\bee\label{diffchain}
\p_{\ga}:=\rho\ff{\p}{\p\theta^{\ga}}\,,\qquad
\bar\p_{\pa}:=\bar\rho\ff{\p}{\p\bar\theta^{\pa}}
 \eee
and  denoting $A^{(p)} \sim (k)^p $ , $A^{(\bar p)} \sim (\bar k)^{\bar p} $, one can see that,
\begin{align}
\p_{\ga}(F^{(p)}*G^{(q)})=\p_{\ga}F^{(p)}*G^{(q)}+(-)^p F^{(p)}*\p_{\ga}
G^{(q)}\,\label{chain},\\\nn
\bar\p_{\pa}(F^{(\bar{p})}*G^{(\bar{q})})=\bar\p_{\pa}F^{(\bar{p})}*G^{(\bar{q})}+(-)^p F^{(\bar{p})}*\bar\p_{\pa}
G^{(\bar{q})}\,.
\end{align}
In \cite{Prokushkin:1998bq,Vasiliev:Rev} it was observed that,
as a consequence of the properties of the deformed oscillator algebra \cite{Vasiliev:1989qh,Vasiliev:1989re}, the operators
  \bee\label{M=SS} \WMO_{AB}:=  \ff{i}{8}\{  \p_A S\,,\p_B S\}_* \q
 (AB)=(\ga \gb)\cup(\pa\pb) \eee
with both of indices $A,B$ being either undotted or dotted
obey the $\mathfrak{sl}(2|\mathbb{C})$ relations
 \bee  \label{MM}
&& [ \!\WMO_{AB} \,,\,  \WMO_{MN} ]_*= \ff{1}{2} \big(\gvep_{AN} \WMO_{ BM}
+\gvep_{AM} \WMO_{ BN}+\gvep_{AN} \WMO_{ BM}
+\gvep_{AM} \WMO_{ BN}\big)\,,  \\ \label{SM}&&
 [\WMO_ {AB}\,,S _C]_*=\half\gep_{AC} S _B+\half\gep_{BC } S _A
  \, ,\eee
that underlies Lorentz covariance of the HS equations (\ref{HS1})-(\ref{HS5}).

  \subsection{Perturbative analysis}
\label{Perturbative solutions}

\subsubsection{Contracting homotopy}
Unfolded equation can be solved perturbatively as follows \cite{Vasiliev:1992av, Vasiliev:Rev}.
At every order, equations to  solve have the form
 \be
\label{fg}
\dr_Z f(Z;Y) = g(Z;Y)\,\q \mathrm{d}_Z:= \theta^A
\frac{\p}{\p Z^A}
\ee
with some $\dr_Z$--closed $g(Z;Y)$.
 These  can be solved, for instance, by the
shifted contracting homotopy $\hmt_{q}$  and cohomology projector $\hhmt_{q }$,  that act as follows
\cite{Didenko:2018fgx, Gelfond:2018vmi}
\be\label{homint0} \hmt_{ q} \phi(Z,Y, \theta) =\int_0^1 \ff{dt }{t } (Z+ q)^A\ff{\p}{\p \theta^A}
 \phi( t  Z-(1-t ) q,t \,\theta)\q\hhmt_{q} \phi(Z,Y, \theta)= \phi(-q,Y,0)\,
\ee
obeying the  resolution of identity
\be\label{newunitres}
\qquad\left\{ \dr_Z\,,\hmt_{q }\right\} +\hhmt_{q }=Id\, .
\ee
At $q=0$, \eq{homint0} yields the Poincar\'e formula (conventional homotopy of \cite{Vasiliev:1992av}).
  \subsubsection{Differential homotopy}
\label{difhom}

Repeated application of   contracting homotopies leads to multiple integrals
$\int  {d t^1 }\int {d t^2 } \ldots$.
All  integration parameters $t_i$  as well as the shift parameters $q_i$  were    interpreted in
 \cite{Vasiliev:2023yzx} as coordinates  $t^a$  with  anticommuting differentials
 $dt^a$ and the total differential
 \be
\label{td} \dr=    \theta^A \ff{\p}{\p Z^A}+ dt^a\ff{\p}{\p t^a}\,. \ee

In these terms, perturbative equations to be solved
acquire  the form \be \label{totg} \dr f(Z,t ,\theta,dt ) = g(Z,t ,\theta,dt
).\ee    The consistency condition
   \be\dr g(Z,t ,\theta,dt )=0\,
   \ee
 is fulfilled as a consequence of the consistency of the full system.
Functions like $f$ and $g$ contain in the integrating measure
factors like
$\theta(t^a)$, $\delta (1-t^a)$   \emph {etc} restricting the $t^a$ integration to a locus $\M$ inside some
hypercube. With such a measure the integration can be formally extended to  $\mathbb{R}^n$.
Physical fields and equations in HS theory
are in $\dr_t$ cohomology supported by the integrals over $t^a$.

  Differential homotopy   is based on the removal of the integrals.
Namely, following \cite{Vasiliev:2023yzx} let \be\label{dfint}
\dr_Z f_{int} = g_{int}\q f_{int} = \int_{\M} f(Z,\theta, t, d t)\q
g_{int} = \int_{\M} g(Z,\theta, t,d t)\,,
\ee
resulting in
\be
\dr_Z f = g +\dr_t  h +g^{weak}\,,
\ee
where $g^{weak}$ does not contribute to the integral because its  degree $\deg g^{weak}$ as a form in the locus $\M$ of $t_i$ differs
from $\dim \M$. Setting $g^{weak} = \dr_Z h - \dr_t f $ (taking into account that
$\deg h=\dim \M-1$ and $\deg f =\dim \M$) and replacing $f\to f+h$ we obtain (\ref{totg}).

Though the integrals are removed from the equations,
 to avoid a sign ambiguity due to (anti)commutativity of
differential forms,   every differential expression
is   accompanied with  integrals.
As in \cite{Vasiliev:2023yzx}, we use notations
\be \int_{\M} f d t^1\ldots d t^k= \int_{t ^1}\ldots \int_{t^k} f :=
 \int_{ t ^1\ldots t ^k} f\, \ee with the convention that $\int_{t ^1\ldots t ^k}$
 is totally antisymmetric
in  $t^a$.
  Integrals   $\int_{t ^1\ldots t ^k}$   can
be written anywhere in the expression for the differential form to be integrated
with the convention
\be \label{MVintconv}
\dr \int_{t ^1\ldots t ^k} = (-1)^k \int_{t^1\ldots t^k} \dr  \,.
\ee

The form of the HS equations within the differential homotopy setup is slightly modified by exact
terms that do not contribute upon integration. Namely, in this setup
 the field  $B$  contains the differentials
$dt^a$ and $\theta^A$ as well as the space-time differentials that enter via the one-form $\go$,
\ie \be \label{cB} \B=\sum_{n\geq 0} B^{(n)} \ee with $B^{(n)}$   containing   $n$ one-forms $\go$
$$
B^{(n)}=\tilde B^n \go^n\, .
$$

  The form of the HS equations in the differential homotopy approach   is \cite{Vasiliev:2023yzx}
\be \label{B} (\dr_x -2i \dr) \B + [\widetilde{\W}\,, \B]_* \cong 0\,, \ee \be \label{dW} (\dr_x -2i\dr) \widetilde{\W} +
\widetilde{\W} * \widetilde{\W} \cong i(\B*(\eta \gga +\bar \eta \bar\gga))\,, \ee
   where \be
\label{cW}
 \widetilde{\W} := \W-\theta^A Z_A \,,
\ee
 while
 the symbol $\cong$ implies {\it weak equality} up to the terms not contributing under
integration.

\subsection{ $\eta$  and $\bar\eta$ deformations}
\label{AnsatzMV}
Here we recall   the Anzatz for
(anti)holomorphic deformation linear in ($\bar \eta$)$\eta$  \cite{Vasiliev:2023yzx}.
Since HS equations  are consistent
with the fields $\W$ and $\B$ valued in any  associative algebra \cite{Vasiliev:1988sa},
different orderings of the fields $\W$ and $\B$
can be considered independently.

  As shown in \cite{Vasiliev:2023yzx},  direct computation within the differential homotopy approach yields the following
 form for the  lowest-order holomorphic deformation linear in $\eta$, where, for the sake of simplicity,
   the integration variables $\rho$ and $\gb$ of
 \cite{Vasiliev:2023yzx}  are set to zero,
 \be
\label{feta}
f_\mu=  \eta\int_{  \tau\gs  }\ls\mu(\tau,\gs ) \dr \Go^2
\Ex  G_l(g(r))\big |_{r=0}\q
\dr \Go^2 := \dr \Go^\ga \dr \Go_\ga\,,
\ee
\be\label{Exp}
\Ex:=   \exp i \big ( \Go_\gb(y^\gb +p^\gb_+ ) -\!\sum_{k\geq j>i\geq 1} p_{i\gb}
p_j^\gb \big)
\q\ee
\bee\label{Omega=}
  \Go_\ga  := \tau z_\ga - (1-\tau)  p_\ga (\gs)    \,,
\eee
\be \label{p+=}
p_{+\ga} = \sum_{i=1}^k p_{i\ga}\q p_{j\ga} = -i \frac{\p}{\p r^j{}^\ga}\q
p_\ga(\gs)=\sum_{i=1}^k p_{i\ga} \gs_i\,,
 \ee
 \be
\label{Gl=} G_l(g):= g_1(r_1)\ldots g_l(r_l) k \,,
\ee
$g_i(y)$ are some functions $C_{j_i}(y)$ or $\go_{n_i}(y)$ of $y_\ga$
(anti-holomorphic variables $\bar y_\dga$, Klein operators
$K=(k,\bk)$ and the antiholomorphic star product $\bar *$ are implicit).
\be\label{fulldifhom}
\dr =d z^\ga \ff{\p}{\p z^\ga} +d\tau\ff{\p}{\p \tau}
 +  d\gs_i \ff{\p}{\p \gs_i}  \,
\ee
with integration parameters $\gs_i$  being some coordinates $t_i$.

The   measure $\mu \dr\Go^2$ may contain some { \it weak}
 terms   that do not contribute under the integration if the number of integrations
 does not match that of respective differentials.

Since $(\dr\Go)^3=0$ as a consequence of the anticommutativity
of $\dr \Omega_\ga$ and two-componentness of the spinor indices $\ga$,
formula (\ref{feta}) has the following remarkable property  \cite{Vasiliev:2023yzx}
\be\label{dGOGO=0} \dr\big(  \dr\Go^2\Ex\big)=0\,.
\ee

 The $\bar \eta$ deformation has the conjugated form. The
 $ \eta\bar\eta$ deformation has the form of the tensor product of the
 $\eta$ and $\bar\eta$ deformations.
In the sequel we use notations of \cite{Vasiliev:2023yzx,Gelfond:2023yzx}
\bee&&\label{nabla}
  l(\gn)=\theta(\nu)\theta(1-\nu)\q
  \tr(\nu)=d \gn \gd(\gn )\,.
 \eee

To simplify formulae in specific calculations   we sometimes   use shorthand notations $\go \go  CC$ instead  of
$\go(r^1,\bar r^1)\go(r^2,\bar r^2)C(r^3,\brr^3) C(r^4,\brr^4)  |_{r^i=\brr^i =0}$  {\it etc}.

\section{Lorentz covariant higher-spin  equations via BRST}
\label{LCHS}
\subsection{General setup}
Consider   "outer'' generators $\gT_{AB}= \gT_{BA}$
with non-zero components $\gT_{\ga\gb}$,  $\gT_{\dga\dgb}$,
that satisfy the $\mathfrak{sl}_2(\mathbb{C})$ relations
  \bee\label{tau}\ls  [\gT_{AB}, \gT_{DC}]_*
=\half( \gep_{AD} \gT_{BC}+ \gep_{AC} \gT_{BD}+
\gep_{BD} \gT_{AC}+ \gep_{BC} \gT_{AD})\,.
 \eee

Associated ghost variables
 $c^{\ga\gb}\,, b^{\ga\gb}\,, \bar{c}^{\pa\pb}\,, \bar{b}^{\pa\pb}
 $
 anticommute with differentials  $\dr x$   and  $\theta$ and
  satisfy the (anti)commutation relations
\bee\label{ghosts}  \{  b_{MN} ,c^{AB}\}_*= \left(\gd_M^A \gd_N^B+\gd_M^B\gd_N^A\right)\q
\{  c^{MN} ,c^{AB}\}_*= \{  b_{MN} ,b_{AB}\}_*=0
\q  [c^{AB},k]=0\,.   \eee

  The finite-dimensional  space $\PP$
  of polynomials of  $\dr x^{\underline{m}}$,  $\theta^A$ and ghosts $c,\,\,b$ \eq{ghosts}
is  endowed with the following $\mathbb{Z}$ grading
\be\pi(c )=\pi(\dr x )=\pi(\theta )=1\,,\,
\pi(b )=-1 .\ee
The fields $\W$ and $\B$ have definite grades  $\pi(\W)=1$, $\pi(B)=0$.
More in detail, decomposing $\PP$ into the direct sum
 of the spaces of three-graded polynomials $P^{n,l,m }$
 in   $\dr x$, $\theta $ and ghosts,
  \be\label{Pgraded} P^{n,l,m
}= \bigcup_{k-s=m}(\dr x)^n (\theta)^l (c)^k (b)^{s}p_{ n,l,k,s}\,,  \ee
\be\label{Wgraded} \W\in \cup_{n+l+m=1}  P^{n,l,m} \q
 \B\in \cup_{n+l+m=0 } P^{n,l,m}.
\ee

Consider equations \eq{hs1}, \eq{hs2}  with the fields $\W$ and $B$  depending in
addition on $c$, $b$ and $\gT$.
These  equations are invariant under the usual HS gauge transformations
\bee\label{gdWB} \gd  \W = [\gvep,\W]_* \q \gd \B = [\gvep, \B]_*
\eee
with grade zero gauge parameters $\gvep$ also dependent on $c$, $b$ and $\gT$, and the additional ones
\bee \label{gdB}\gd \B = \{\gf, \W\}_*\q \gd
\W = \gf \,*( i\eta  \gga+i\bar{\eta} \bar{\gga})\, \eee
with grade $-1$ gauge parameters $\gf$
 as can be easily checked taking into account that
$\gga$ and $\bar\gga$ (\ref{gamma}) are central.
  Note that   transformations \eq{gdB} were
absent in the original equations \eq{hs1}, \eq{hs2} where $\B$ was a ghost-independent zero-form as there was no room for the
grade $-1$ gauge parameters
$\gf$ in the absence of the ghosts $b_{\ga\gb}$ and $  \bar{b}_{\dga\dgb}$.

\subsection{Field decomposition}
\label{EVALu}

To define a proper BRST operator we consider
the following mutually commuting "inner" \\ $\mathfrak{sl}(2|\mathbb{C})$-generators
 $L^Y _{AB}$, $L^Z_{AB}$\,, $L^{\theta} _{AB}$ and $L^{ghost} _{AB}$ with $ A B =(\ga\gb,\,\,\,\pa\pb)$
   \bee&&  \label{L=}
L^Y_{AB}f=\half \left(  Y_{A}\ff{\p}{\p Y^B}\! +\!Y_{B}\ff{\p}{\p Y^A}\right)f
\equiv \!-\!\ff i4  [ Y_A Y_B \,,f]_*
\q\qquad\\&&  \label{LZ=}
L^Z_{AB}f=\half \left(Z_{A}\ff{\p}{\p Z^B} +\!Z_{B}\ff{\p}{\p Z^A} \right)f
\equiv  \ff i4 [ \!Z_AZ_B \,,f]_*
\q\qquad\\ && \label{Lteta=}
L^\theta_{AB}f= \theta_{(B}\ff{\p}{\p\theta^{A)}}= \half
 \left(\theta_{\ga}\ff{\p}{\p\theta^\gb}+\theta_{\gb}\ff{\p}{\p\theta^\ga}\right)f+
\half \left(\bar\theta_{\pb }\ff{\p}{\p\bar\theta^\pa}+\bar\theta_{\pa}\ff{\p}{\p\bar\theta^\pb}\right)
\q\\  &&\label{Lghostt=}
L^{ghost}_{AB}f= \ff{1}{2} \big\{ c_{A}{}^{C} b_ {CB}+c_{B}{}^{C} b_ {CA}
\,,\,f\big\}\q\eee that commute with $\gT$ \eq{tau} \bee\label{tau L comm}
[\gT_{AB}, L^Y]_*=[\gT_{AB}, L^Z]_*= [\gT_{AB}, L^\theta]_*=[\gT_{AB}, L^{ghost}]_*= 0 .
 \eee

 Note, that  \bee&& \label{Ltot=}L^{Tot}= L^{Y}+L^{Z}+L^\theta+L^{ghost}+\gT \eee
is also an $\mathfrak{sl}(2|\mathbb{C})$-generator.
   The BRST  operator  relevant to the furher analysis is
 \bee\label{Q=} Q=c^{AB}(\gT_ {AB} +L^\theta_ {AB}+
 L^Z_ {AB}
 )-  \half
 c^{A}{}_{D}c^{D B}b_{AB}\,.
 \eee
 It obeys $Q^2=0$   by virtue of  \eq{tau}, \eq{Lteta=}, \eq{tau L comm}, and
  \bee \label{bQ} \{ b_{AB}, Q\}=
 2(\gT +L^\theta + L^Z + L^{ghost})_ {AB}.
 \eee

Following the idea of  \cite {Didenko:2017qik},
in addition to the Lorentz
connection  $  \go^L $ we introduce    a new Lorentz
connection one-form  \be\label{hatgo} \hat \go=\hat{\go}^{AB}(x) L^{Tot}_{AB}\ee
and a two-form $U^{AB}(x)$ that will play a role of its curvature.
Note  that $\hat{\go}^{AB}(x)$ and $U^{AB}(x)$ are independent of
$Y,Z,c,b,\theta,\gT$.

 Consider the    decomposition
\bee \label{GW=W +U+}  \ls\W &\!\!:=\!\!&
\hat{\go}^{AB}(x) L^{Tot}_{AB}+\half U^{AB}(x) b_{AB}  +\W' +Q
 \,\q\eee
where
 $\W'( Y,Z ;c,b ; \theta,\gT |x,\dr x)$ is independent of the  terms like
  \bee&& \label{ISKLW}
   G^{AB}(x) b_{AB} \q G^{AB}(x ) \gT_{AB}\q G_{CD}^{AB}(x )c^{CD} \gT_{AB}
   .\eee
 Substitution of \eq{GW=W +U+}   into \eq{hs1}, \eq{hs2}
 yields
by virtue of  \eq{tau}, \eq{ghosts}, \eq{L=}-\eq{bQ}\bee
  \ls&& \label{GdxW=}
  \D^L \W'+ \half(\dr_x U^{AB}+2
 \hat{\go} _{D}{}^{A}  U^{D B} )b_{AB}  \!+\! \W'* \W'\!+\! \{\W'\,,Q\}_*
  +\\ \nn &&
+ (U^{AB} +d_x  \hat{\go}^{AB} \!+\!   \hat{\go}_{C} ^{ A}\hat{\go}^{B}{}^C )
 L^{Tot} _{AB}% (L_{AB}^\theta+L^{ghost}_{AB}+\gT_{AB}+\ff i4 Z_{A }Z_{B})
 +\ff i4   U^{AB} Y_{A }Y_{B}
   \!+\!\half \{U^{AB}  b_{AB}\,, \W'\}_*  +   \\  \ls&&\nn
\rule{0pt}{14pt} +i\theta_{\ga}\wedge \theta^{\ga}(1+\eta \B*k\gk)+
i\bar\theta_{\pa}\wedge \bar\theta^{\pa}(1+\bar\eta \B*\bar k\bar \gk)
   =0\q
\\ &&\label{G dxB}
\rule{0pt}{16pt}
 \D^L \B   +[U^{AB}  b_{AB}+\W'+Q\,, \B ]_*  =0\,\q  \eee where
\bee&&
 \label{DL=} \D^L f:=\dr_x f+\hat{\go}^{AB} [ L^{Tot}_{AB} ,f]_*.\eee

Note that since  $\W' $ is free of the terms
 \eq{ISKLW},
the condition that the $\gT-$dependent term on
the \lhs of \eq{GdxW=} is zero implies
 \bee\label{U=}   U^{AB} +d_x  \hat{\go}^{AB} \!+\!   \hat{\go}_{C} ^{ A}\hat{\go}^{B}{}^C =0
.\eee
Since consistency of \eq{U=} implies the Bianchi identity for the curvature two-form
$U^{AB}$,
 \bee\label{dU=}\dr_x U^{AB}+2
 \hat{\go} _{D}{}^{A}  U^{D B} =0,
  \eee
  \eq {GdxW=} amounts to
\bee
  \ls&& \label{GdxW=gT}
  \D^L \W'+      \W'* \W'+\{\W'\,,Q\}_*
  +\ff i4 U^{AB}Y_{A }Y_{B}
   +\half \{U^{AB}  b_{AB}\,, \W'\}_*  +   \\  \ls&&\nn
\rule{0pt}{14pt} +i\theta_{\ga}\wedge \theta^{\ga}(1+\eta \B*k\gk)+
i\bar\theta_{\pa}\wedge \bar\theta^{\pa}(1+\bar\eta \B*\bar k\bar \gk)
   =0\,.
\eee
 \subsection{Gauge symmetries }

 Substitution of $\W$ \eq{GW=W +U+} into  \eq{gdWB}, \eq{gdB} with  $\gvep=\gvep'+  \gx^{AB}(x) b_{AB}\,,$ where $\gvep'$ is
  independent of terms like $f^{AB}(x) b_{AB}$, yields
   \bee \label{SUBS}&&\gd\hat{\go}^{AB}(x) L^{Tot}_{AB}+\half\gd U^{AB}(x) b_{AB}
  +\gd \W'( Y,Z ;c,b ; \theta,\gT |x,\dr x)
 =\\ \nn&&= 2\gx^{AB}(x) (\gT +L^\theta +
 L^Z +  L^{ghost} )_{AB}- ( \dr_x  \gx^{AB} +
2 \hat{\go} _{D}{}^{A}  \gx^{D B}) b_{AB}+
\\&&+ [\gx^{AB}(x) b_{AB},\W']-\D^L(\gvep')+\big[\gvep',\, \W'{} +Q+\half U^{AB} b_{AB}\big]_*\nn\,,\eee
\bee \nn&&  \gd\B = \big[\gvep'+  \gx^{AB}(x) b_{AB}, \B\big]_*
+\D^L(\gf)+\{\gf\,, \W'{} +Q+\half U^{AB} b_{AB} \}_*.\eee
Hence  equations \eq{GdxW=}, \eq{G dxB}
are invariant under the following gauge transformations:
  \bee \label{gdksi} &&\gd \hat{\go}^{AB}=2\gx^{AB}\,,
  \\   \label{gdUW}&&   \gd U^{AB}= -( \dr_x  \gx^{AB} +
2 \hat{\go} _{D}{}^{A}  \gx^{D B}) \,,\\
    \label{gdUWpr}&&\gd \W'{}=    \gf  *\big( i\eta  \gga+i\bar{\eta} \bar{\gga}\big)\,
+  \gx^{AB}\big ( - 2L^Y+ \{b_{AB} ,\W'{}\} \big)
-\D^L(\gvep')+\\ \nn&&+\big[\gvep',\,
 \W'{} +Q +\half  U^{AB} b_{AB}\big]_* \q
\\\label{gdBB}&&  \gd\B = \big[\gvep'+  \gx^{AB}(x) b_{AB}, \B\big]_*
+\D^L(\gf)+\{\gf,  \W'{} +Q+\half U^{AB} b_{AB} \}_*.\eee

Note that the gauge transformations with the parameter $ \xi^{AB}$,
%Symmetries
 \bee \label{SHTUK}
&&\gd  \hat{\go} _{AB}=2\xi {}_{AB}
%+  d_x {\gvep}_{AB}
 \q\\ \nn
&&\gd  U^{AB}=- d_x  \xi^{AB} - \hat{\go}^{A}_{D } \xi ^{DB} -\hat{\go}^{B}_{D
} \xi ^{DA}\q\\ &&\nn \gd \W' =  \gx^{AB}\big ( - 2L^Y_{AB}+ \{b_{AB} ,\W'{}\} \big) \q\\
\nn&&\gd\B = \big[  \gx^{AB}(x) b_{AB}, \B\big]_*  \eee
form  a  Stueckelberg  symmetry
  allowing  to gauge away either $\hat{\go}_{AB } $ or  $\go^L_{AB}$. In the  gauge $\hat{\go}_{AB }=0 $
 the original formulation of the HS theory is recovered, while in the gauge $\go^L_{AB }=0 $,
 that with manifest Lorentz symmetry results.

 \subsection{Vacuum solution }
\label{VacQ}
Let \bee B =B_0,\quad \W'_0 =W_0+S_0,\quad U=U_0,\quad\hat{\go}  =\hat{\go}_0{},\quad B_0=0 ,\quad
  \label{S0gh}  S_0=\theta^A Z_{A} +Q \eee
  with $Q$ \eq{Q=} and the
fields $W_0$,  $U_0$   and  $\hat{\go}_0{}^{AB}$   defined below.
According to \eq{U=}\,, \bee\label{U=0}
 U_0{}^{AB} +d_x  \hat{\go}_0{}^{AB} \!+\!   \hat{\go}_0{} ^{ A}{}_{C}\hat{\go}_0 {}^{B}{}^C =0
\,.\eee
Since
\be \{Q,\theta^A Z_{A}\}_*= 0 \q S_0*S_0=i\theta^{A}\theta_{A}
  \q \label{dZeq}
[\theta^A Z_{A}\,,\cdot]_* = -2i \mathrm{d}_Z\,,
\ee
 plugging $W_0$ into  \eq{GdxW=gT} one
finds that
  \bee\label{newdW0}    \D_0^L    W_0        +
         W_0  * W_0  \!+\!   \ff i4  U_0{}^{AB}( x)  Y_AY_B
+ \{(-2i\dr_Z{}+Q), W_0  \} =0  \,.
\eee

From \eq{dZeq} it follows that  $(-2i\dr_Z{}+Q)^2=0$ in the graded commutators.
On the  space $P^{n,l,m }$, $\dr_Z{} $ and $Q$ act as follows
 \bee \dr_z
P^{n,l,m}\subset P^{n,l+1,m}\q Q P^{n,l,m}\subset P^{n,l,m+1}\,.
 \eee Therefore,   from \eq{newdW0}  it follows
 that the space-time one-form $W_0$ satisfies
 \bee  \label{dz+Q} [(-2i\dr_Z{}+Q), W _0  ]_\pm&=& 0\,.
  \eee
  Using the spectral sequence analysis, one can make sure that
  \be \label{W_0=} W _0  =(-2i\dr_Z{}+Q) G  +
\go(x, Y, c,b, \gT )\q Q \go=0\,,
\ee
where the term with $G $ is pure gauge and can be gauge fixed to zero.
 Any $Q$-exact $\go(x, Y, c,b, \gT )$  also belongs to $G$.
In accordance with the  Whitehead Lemma
\cite{Jakobson}\,,
  $Q$ cohomology  of simple finite-dimensional
Lie algebras with coefficients in finite-dimensional representations
is trivial in all degrees except for zero
 where it is equal to the ground field hence being
represented by a constant in $c,b$ and $\gT $, \ie by
\be
\label{convent go}
 \go=\go(x, Y  )\,.\ee
This conclusion is the same as in the original  HS equations of \cite{Vasiliev:1992av} where  $\go(x, Y  )$ is the
generating function for HS gauge fields.

 From  \eq{newdW0}  one finds
 that $\go$ satisfies the
 equation \bee&& \label{Uvacua1} \D^L\go =-   \go*\go - \ff i4\left(U_0^{\ga\gb}y_{\ga } y_{ \gb}+
  \bar U_0^{\pa\pb}  \by_{\pa } \by_{ \pb}
   \right) + \mbox{higher-order terms}\,.\eee
For $\go$  demanded to be free of the Lorentz connection,  the respective terms of
$\go*\go$ can be compensated by $U_0$.\footnote{If $\go$ belongs to a  matrix algebra\,,  the Lorentz connection
$\go $  and $U_0$ are proportional to the unit matrix.}  To find
$U_0$ we set to zero components of $  \go *\go  $
   bilinear either in $y$ or in    $\by$,
\bee\label{Tr0} && \ls\ls\ls\ff{\p^2  \big(
  \go *\go  \big)}{\p y^\gn\p y^\mu} |_{ y =\by =0}
 +
    \ff i2 U_0{}_{\gn\mu}=0,\quad \ff{\p^2  \big(
  \go *\go  \big)}{\p \by^\pa\p \by^\pb}|_{ y =\by =0}
 +
    \ff i2 \bar  U_0{}_{\pa\pb}=0\,.\quad \eee

For instance consider   vacuum $\go=\go_0$ free from the Lorentz connection,
 \be \label{nR2w} \go_0=\ff 12  \lambda h^{\ga}{}^{\pb}y_\ga\by_\pb \,. \ee
From \eq{Tr0}   it follows
\bee &&  \label{Uvacua1L}
U_0{}^{\ga\gb}=- \lambda^2\,h^{\ga}{}^\pa h^\gb{}_\pa\q
\bar U_0{}^{\pa\pb}=-\lambda^2\, h^{\ga}{}^\pa h_{\ga}{}^{\pb} .
 \eee
  The \rhs of \eq{Uvacua1} with $U_0$  \eq{Uvacua1L}  does not contain the Lorentz connection component of $\go$.
Instead \eq{U=0} yields equations of $AdS_4$ in the form
 \be\label{vacgo0L} d_x  \hat{\go}_0^{\ga\gb} +   \hat{\go}_0{}_{\gga}{}^{\ga}\hat{\go}_0{}^{\gb}{}^\gga   =
 \lambda^2 h^{\ga}{}^\pa  h^\gb{}_\pa\,\q
d_x  \hat{\go}_0^{\pa\pb} + \hat{\go}_0{}_{\dgga}{}^{\pa}\hat{\go}_0{}^{\pb}{}^\dgga   =
 \lambda^2
h^{\ga}{}^\pa h_{\ga}{}^{\pb}
\ee
with the vacuum Lorentz connection $\hat{\go}_0$.

\subsection{Perturbative analysis}
\label{Perturbative analysis}
The idea of the proposed approach is to use the  Stueckelberg   symmetry \eq{SHTUK} to
gauge away the Lorentz connection parts  of $  \go^L$  proportional to $y_\ga y_\ga$ or
$\bar y_\dga \bar y_\dga$  in favour of $\hat \go$ \eq{hatgo}. On the one hand, this
will make all derivatives   Lorentz covariant with $\hat \go$ entering
solely via the Lorentz covariant derivative $\D^L$ \eq{DL=}. On the other, the condition that
$\go(Y)$ is free of the terms proportional to $y_\ga y_\ga$ and
$\bar y_\dga \bar y_\dga$ will express $U_{\ga\gb}$ and $\bar U_{\dga\dgb}$ via dynamical
massless fields thus inducing nonlinear corrections to the Lorentz covariant form of the HS field equations.
The result is analogous to the contribution of
the tensor $R$ in the approach of \cite{Didenko:2017qik} with the difference that   the latter was found
within a specific homotopy scheme from the consistency conditions while in our approach $U_{\ga\gb}$ and $\bar U_{\dga\dgb}$
are reconstructed order by order from the system of equations  consistent in any  homotopy scheme.

\subsubsection{First order}

Let us illustrate this computation scheme by the linearized analysis. To this end we set
\be\label{lineardata}\B = B_1\q\W' =\go+S_0+W_1+S_1\q U=U_0+U_1\q\hat{\go}  =\hat{\go}_0+\hat{\go}_1 \ee with
$S_0 $\eq{S0gh}, $U_0 $\eq{Uvacua1L} and $ \go$ \eq{convent go}. We assume that
$S_1$ is free of $\dr x$ and $W_1|_{\dr x=0}=0$.

Then
Eqs.~\eq{U=}-\eq{GdxW=gT} and \eq{G dxB} yield
\bee\label{U1=}  && U_1^{AB}+d_x  \hat{\go}_1{}^{AB}+ \{ \hat{\go}_1{}_{C} ^{ A}\,,
\hat{\go}_0{}^{B}{}^C \}=0\,,\\&&\label{dU1=}\dr_x U_1^{AB}+2
 \hat{\go}_0{}_{D}{}^{A}  U_1{}^{D B}+2
 \hat{\go}_1{} _{D}{}^{A}  U_0{}^{D B} =0\,,\eee
\bee   \label{EXG W}
  \ls&&\D^L  (\go+W_1) + \go\,* \go +\ff i4 (U_0+U_1)^{AB}Y_{A }Y_{B} +
    \{ W{}_1{}\,, \go\}_* +\{ W{}_1{} \,, S{}_1{}\}_*
    \\ \nn &&
   + \half\{U{}_0{}^{AB}  b_{AB}\,, (W{}_1{}+S{}_1{})\}_*
  +\D^L  S{}_1{}+    \{ W{}_1{} \,,(-2i\dr_Z{}+Q) \} +    \{ \go{} \,, S_1\}_* +
\\
      \ls&& \nn
      + \{S{}_1{} \,,(-2i\dr_Z{}+Q){}\}_*
  \rule{0pt}{14pt} +i\eta\theta_{\ga}\wedge \theta^{\ga}   B{}_1{} *k\gk +
i  \bar\eta \bar\theta_{\pa}\wedge \bar\theta^{\pa} B{}_1{} *\bar k\bar \gk
   =0\,,
\eee
 \bee&&
 \label{EdxdB} \D^L  B{}_1     +[ \go  \,,B{}_1   ]_*
 +\half[U{}_0{}^{AB}  b_{AB} ,B{}_1{}  ]_* +[(-2i\dr_Z{}+Q) \,,B{}_1{}  ]_*   =0\, .\eee

 Analogously to the analysis of formulae \eq{newdW0}-\eq{convent go},   from \eq{EdxdB} it follows
 \bee \label{EBS+QB}  [(-2i\dr_Z{}+Q), B_1 ]_*  =0\,\eee
and hence
\be\label{convent C} B_1=C(x, Y )+(-2i\dr_Z{}+Q)G.\ee
 The $G$-dependent term can be gauge fixed to zero by virtue of the gauge symmetry \eq{gdBB}. Then \eq{EdxdB} yields
 the standard linearised HS equation
 \bee &&
  \label{EdxdC} \D^L C  +[ \go{} ,C  ]_*=0.\eee
 Since   $S_1$ does not contain $dx$,    \eq{EXG W} yields
 \bee \label{EBS+Q}  \{ -2i\dr_Z{}+Q\,, S_1 \}_*
  \rule{0pt}{14pt} =-i \eta \theta_{\ga}\wedge \theta^{\ga}\, C *k\gk -
i \bar\eta \bar\theta_{\pa}\wedge \bar\theta^{\pa} \, C *\bar k\bar \gk
  \,.\eee
Note that, since $Q$ acts    on $Z$ and $\theta$  non-trivially,
$S_1 $ should depend  on $Z$, $\theta$ and ghosts.
Let us decompose it into the ghost-independent and ghost-dependent parts $S'$ and $S''$, respectively,
  \bee
\label{decomp S}
S_1:=S'_1{} + S''_1\q S_1'= S'_1{} {}_\ga (x, Y,Z)\theta^\ga+\bar S'_1{} {}_\pa (x, Y,Z)\bar\theta^\pa \,.
\eee
Suppose the ghost independent component to obey
  \bee \label{convent S}
 -2i\dr_Z{}   S'_1{} {}
 \rule{0pt}{14pt} +i \eta \theta_{\ga}\wedge \theta^{\ga}\, C *k\gk +
i \bar\eta \bar\theta_{\pa}\wedge \bar\theta^{\pa} \, C *\bar k\bar \gk
   =0\,. \eee
Solutions to \eq{convent S} are well known. In notations of Section
\ref{AnsatzMV}, equation \eq{convent S}  can be solved in the form \cite{Vasiliev:1992av, Vasiliev:2023yzx}
\bee\label{S1withw}
&&   S '_1  = -\ff{\eta}{2 }\int_{\gt  } l(\gt)
  (\dr\wmv )^2\EE(\wmv ) C * k-\ff{\bar{\eta}}{2 }\int_{\bgt  } l(\bgt)
(\dr   \bwmv )^2   \bEE(\bwmv) C * \bk\q
 \\\label{wS1withw}
&& \wmv^\ga= \gt z^\ga \q  \bwmv^\pa=\bgt \bz^\pa \,.\eee
 Note that
 \be\label{dGO2}\dr   \wmv= d\gt z^\ga+\gt\theta^\ga\quad\Rightarrow\quad(\dr   \wmv )^2\cong 2\gt \dr\gt z^\ga \theta_\ga.\ee
From \eq{EBS+Q} due to \eq{Lteta=}, \eq{Q=} it follows that
\bee \label{EBS+Q-}   \{(-2i\dr_Z{}+Q), S''_1{}  \}_* +c^{\ga\gb} S'_1{} {}_\ga (x, Y,Z)\theta_\gb+
\bar{c}^{\pa\pb} \bar S'_1{} {}_\pa (x, Y,Z)\bar\theta_\pb   =0
 \,.\eee
 Using perturbative decompositions of formulae \eq{MM}, \eq{SM}
one can straightforwardly make sure  that    \eq {EBS+Q-} admits a  solution  \be\label{S''}
 S''_1=\ff{ i}{2}   z_\ga\,   S'_1{} {}_\gb c^{\ga\gb}+
\ff{ i}{2}    \bz_\pa\,\bar S'_1{} {}_\pb  \bar{c}^{\pa\pb}+\widetilde{S}''_1\ee with
$S'_1 $
  satisfying \eq{convent S}, $S'_1{}_A=\p_{\theta^A}S'_1$
and $\widetilde{S}''_1{} $ satisfying
$ \{(-2i\dr_Z{}+Q), \widetilde{S}''_1{}  \}_*=0$ .
The latter implies
$\widetilde{S}_1''{} = (-2i\dr_Z{}+Q)F$  for some $F$, that
   allows the gauge with ${\widetilde{S}_1''{} }{}= 0$.

 Hence \be \label{S'+S''} S_1=S'_1{} +\ff{i}{2}   z_\ga\,   S'_1{} {}_\gb c^{\ga\gb}+
\ff{i}{2}    \bz_\pa\,\bar S'_1{} {}_\pb  \bar{c}^{\pa\pb} \ee
with $S'_1$ \eq{convent S} solves \eq{EBS+Q}.

  Note, that perturbatively, the $\mathfrak{sl}(\mathbb{C},2)$-generator
$\WMO_{AB}$ \eq{M=SS} has the form $\WMO_{AB}= \WMO_0{}_{AB}+ \WMO_1{}_{AB}+\ldots$, where
 \bee\label{M01}
\WMO_0{}_{AB} =\ff{i}{4}   Z_A   Z_B  \q
\WMO_1{}_{AB} =\ff{i}{4}     Z_A   S'{}_B +\ff{i}{4}    Z_B   S'{}_A .\eee
Hence from \eq{S''} it follows  that,     in agreement with \cite{Didenko:2017qik},
\bee
\label{M=SSP}    S''_1{}  \equiv
\WMO_1{}_{AB}c^{AB}\,.\eee

By virtue of \eq{EBS+Q}--\eq{convent S} and \eq{EBS+Q-}, Eq.\eq{EXG W} yields
 \bee \label{EXG WCtilS}
  \ls&&\D^L  (\go+W_1) + \go\,* \go +\ff i4 (U_0+U_1)^{AB}Y_{A }Y_{B} +         \{ W{}_1{}\,, \go\}_*
      \\ \nn &&
   + \half\{U{}_0{}^{AB}  b_{AB}\,, (W{}_1{}+S{}_1{})\}_*
  +\D^L  S{}_1{} +    \{ \go{} \,, S_1\}_*+    \{ W{}_1{} \,,(-2i\dr_Z{}+Q) \}
    =0\,.
\eee
In the sector of space-time one-forms this gives
\bee  \label{DW1=}\D^L   S {}_1{}
+    \{ \go{} \,, S _1  \}_* +    \{ W_1{}  \,,(-2i\dr_Z{}+Q) \}
=0\,.\eee
To find $W _1$ let us decompose $W _1{}=  {W_1'{} } + {W_1''{} } $ with ${W_1'{} }$ obeying
\bee\label{dzW1} {W_1'{} }=W_1'{} (x,Y,Z) :\qquad  -2i\dr_Z{}{W_1'{} }
=-\D^L   S'_1{}  -   \{ \go{} \,, S'_1{}   \}_*
 . \eee
  Note that the Lorentz covariance of the expressions \eq{Exp}, \eq{Omega=} and the conjugated implies
 that  by virtue of \eq{EdxdC}  $S '_1$  \eq{S1withw}   satisfies \be
    \D^L S '_1 (C) = - S '_1 (\D^L C  )=   S '_1 ( [ \go{} ,C  ]_*).\ee As a result,
   \eq{dzW1} acquires   the form
\bee\label{dzW1=}   -2i\dr_Z{}{W_1'{} }
=    - S_1'( [ \go{} ,C  ])  -   \{ \go{} \,, S'_1{}   \}_*\,.
   \eee
   Since $S_1'$ coincides with $S_1$  of \cite{Vasiliev:1992av}
 the  \rhs of    \eq{dzW1=} coincides with that of the equation for $W_1$ of \cite{Vasiliev:1992av}.
 Hence
one can choose a solution coinciding with the conventional $W_1$  originally found in \cite{Vasiliev:1992av} on AdS background
and  then in \cite{Didenko:2018fgx} on an arbitrary zero-curvature HS background.
In notation of Section \ref{AnsatzMV} this yields
 \bee\label{W1'til}\ls {W_1'{} }{}&=& {W_1'{} }{} \,|_{\go C} +{W_1'{} }{}\,|_{C\go }\,,
\eee\bee\label{W1goC}
\ls {W_1'{} }{} \,|_{\go C} &=& \ff{ i\eta}{4 }
\!\!\int\limits_{ \gt,   \gs  } l(\gs)    l(\gt)  (\dr\wmv )^2
 \EE(\wmv)  \go  C  *k
+
\ff{i\bar\eta}{4 }\!\!\int\limits_{  { \bgt},  \gs  } l (\gs)
 l(\bgt)    (\dr\bwmv )^2  \bEE(\bwmv)  \go  C   *\bk
\q  \\
  \label{wW1goC}&&\wmv^\ga =\gt z^\ga-(1-\gt) (- \gs  p_\go  )^\ga  \q
 \bwmv^\pa =\bgt \bz^\pa-(1-\bgt)(- \gs  \bp_\go  )^\pa
 \,\q
\\\label{W1Cgo}
  {W_1'{} }{}\,|_{C\go } &= &\ff{\eta}{4i }
\!\int\limits_{   \gt  \gs  } l (\gs) l(\gt)    (\dr\wmv )^2  \EE(\wmv)     C   \go *k
+
 \ff{ \bar\eta}{4i }
   \int\limits_{  \bgt,  \gs } l(\gs) l(\bgt)  (\dr\bwmv )^2\bEE(\bwmv) C  \go    *\bk
\q  \\
 \label{wW1Cgo}&&\wmv^\ga =
\gt z^\ga-(1-\gt) (\gs   p_\go  )^\ga
  \q  \bwmv^\pa =\bgt \bz^\pa-(1-\bgt)
 (\gs \bp_\go  )^\pa
 \,.\eee
 By virtue of  \eq{dzW1} from  \eq{DW1=} it follows \bee  \label{DS''}\D^L   S''_1{} {} {}
+    \{ \go{} \,, S''_1{}   \}_* + \{  W_1'{}   \,,Q\}_*+   \{ W_1''{} {}  \,,(-2i\dr_Z{}+Q) \}_*
=0\,.\eee

Using \eq{MM}, \eq{SM}, \eq{S''} and \eq{dzW1},  one can make sure that   the following equation holds true
 \bee\label{DLW_1''{} }\D^L   S''_1{} {} {}
+    \{ \go{} \,, S''_1{}   \}_* + \{  W_1'{}   \,,Q\}_*=0 \,. \eee
 Then from \eq{DS''}    it follows
   $ \{ {W_1''{} }{}  \,,(-2i\dr_Z{}+Q) \}
 =0 .$
As before, this implies
\be\label{convent W2} W_1''{} =\omega_1 (x, Y )+(-2i\dr_Z{}+Q)G_1,\ee
for some $\go_1$ and $G_1$. This allows the gauge with ${W_1''{} }{}= \omega_1 (x,Y)$. Renaming $\omega +\omega_1$ by a redefined
$\omega$  we can set
   \be\label{W_1''{} til} {W_1''{} }{}=0 .\ee
As a result, by virtue of \eq{decomp S}, \eq{M=SSP},    \eq{DW1=}, \eq{W_1''{} til} and   taking into account  \eq{ISKLW}
one finds from \eq{EXG WCtilS}
\bee  \label{dLWCW1} \ls&&\D^L   \go  =-\big( \go\,* \go+\D^L{W_1 {} }+ \ff i4 (U_0+U_1)^{AB}Y_{A }Y_{B}     +     \{ {W_1 {} }{} \,, \go\}_*
 +   U{}_0{}^{AB}  M_1{}_{AB}       \big) \,.
\eee
Since $\go$ was demanded to be free from Lorentz connection, taking into account \eq{Tr0}
the respective terms of \be R_1:= \D^L   W_1 {}  +     \{  W_1  \,, \go\}_*
   +  U{}_0{}^{AB}  M_1{}_{AB}\ee
 bilinear in $y$ and  in  $\by$  have to be compensated by  $U_1$,
\bee\label{Tr1} && \ls\ls\ls\ff{\p^2 R_1 }{\p y^\gn\p y^\mu} |_{ y =\by=0}
 +
    \ff i2 U_1{}_{\gn\mu}=0,\quad \ff{\p^2 R_1 }{\p \by^\pa\p \by^\pb}|_{ y=\by=0}
 +
    \ff i2 U_1{}_{\pa\pb}=0\,.\quad \eee
The Lorentz connection $\hat{\go}_1$ satisfies \eq{U1=} with $U_1$ \eq{Tr1}.

\subsubsection{To higher orders}
The analysis of the two first perturbative orders suggests
 that, analogously to \eq{lineardata}, \eq{decomp S}, it is convenient to decompose
\be \W' =\go+S_0+W+S\,,\ee
where $S $ is free of $\dr x$ and $W |_{\dr x=0}=0$,  $S_0 $ \eq{S0gh}
and $ \go$ \eq{convent go}.
Then from  \eq{EXG W} it follows
\bee   \label{DLWfull}
  \ls&&\D^L  (\go+W) + (\go+W)\,*(\go+W) +\ff i4  U ^{AB}Y_{A }Y_{B} +
     \half\{  U  ^{AB}
    b_{AB}\,, (W {}+S {})\}_* =0\q\\ \label{DLSfull}
&&   \D^L  S {}  +\{ W {} \,, S {}\}_* +    \{ W {} \,,(-2i\dr_Z{}+Q) \} +    \{ \go{} \,, S \}_*
 =0\q
\\      \ls&& \label{dSfull}
        \{S{} \,,(-2i\dr_Z{}+Q){}\}_*+ S\,* S
  \rule{0pt}{14pt} +i\eta\theta_{\ga}\wedge \theta^{\ga}   B{}  *k\gk +
i  \bar\eta \bar\theta_{\pa}\wedge \bar\theta^{\pa} B{}  *\bar k\bar \gk
   =0\,.
\eee
Let
 \bee\label{decomp S full}
\ls&&S :=S' + S''
\q S'= S' {}_\ga (x, Y,Z)\theta^\ga+\bar S' {}_\pa (x, Y,Z)\bar\theta^\pa\,,
\eee
where $S'$ is defined to obey
       \bee &&\label{S'eq}-2i\dr_Z S'{}=- S'\,* S'
  \rule{0pt}{14pt} -i\eta\theta_{\ga}\wedge \theta^{\ga}   B{}  *k\gk -
i  \bar\eta \bar\theta_{\pa}\wedge \bar\theta^{\pa} B{}  *\bar k\bar \gk
\,.\eee
Then \eq{dSfull} yields\bee \ls&& \label{S''eq}
        \{S''{} \,,(-2i\dr_Z{}+Q){}\}_* +\{S' {} \,, Q{}\}_*+ \{S'\,, S''\}_*+S''\,* S''
   =0\,.
\eee
Note that the \rhs of \eq{S'eq} coincides with that of the respective equation on $S-S_0$ of \cite{Vasiliev:1992av}.
   Then $\WMO_{AB}$ \eq{M=SS} can be rewritten as
\bee\label{M=SS'=} \WMO_{AB}= \WMO'_{AB}+\ff{i}{8}\{  Z_A  \,,Z_B  \}_*
\,,\quad\WMO'_{AB}=  \ff{i}{8}\big(\{  \p_A S'\,,\p_B S'\}_*+ \{  Z_A  \,,\p_B S'\}_* +  \{  \p_A S'\,,Z_B  \}_*\big).\eee
From \eq{MM},  \eq{SM}
it follows that
 $ S''{}=\WMO'_{AB}c^{AB}$
solves \eq{S''eq}.

Let us stress that the decomposition of $S$ into $S'$ and $S''$ presented above is available
due to the specific form of the HS equations \eq{HS4} and hence \eq{S'eq} related to the deformed oscillator algebra which still plays a fundamental role in the presented construction of the HS theory. The  LCHS equations can be analysed analogously
at  all higher orders. However, straightforward application of this method
   is rather involved even at the linearized level. Now we are in a position to show that
 using differential homotopy approach  the same results  can be reached  in a much more elegant way.

 \section{Lorentz covariantization within differential homotopy}
\label{LCDifHom}
\subsection{General setup}
In the BRST-extended model we extend the differential $\dr$
 \eq{td} to
 \be
\label{tdgh2} \dr_Q=   \dr_Z   +dt^a\ff{\p}{\p t^a}-\ff{1}{2i} Q
 \ee with
  $Q$ \eq{Q=}. Clearly, $  \dr_Q^2 =0 $.
   Proceeding analogously to
Section \ref{difhom}  one can see that within the differential homotopy approach,
    the LCHS  equations \eq{GdxW=gT} and \eq{G dxB}  acquire the form
 \bee
  \ls&& \label{GdxW=gT==}
 ( \D^L-2i\dr_Q)   \W'+      \W'* \W'+
  \ff i4 U^{AB}Y_{A }Y_{B}
   +\half \{U^{AB}  b_{AB}\,, \W'\}_*  +   \\  \ls&&\nn
    +i\eta\theta_{\ga}\wedge \theta^{\ga}  \B*k\gk +
i\bar\eta\bar\theta_{\pa}\wedge \bar\theta^{\pa}  \B*\bar k\bar \gk
   \cong 0\,,
\\\ls&&\label{G dxB=Dif} (\D^L -2i\dr_Q  ) \B
+[\W\,, \B ]_*  \cong 0\, \rule{0pt}{17pt} \eee
   with  $\D^L$  \eq{DL=} and
   \be
\label{cWprim}
 \W := \W'- 2i\dr_Z  +Q  \,.
\ee
  As mentioned in     Section \ref{AnsatzMV}, to solve HS equations  within differential homotopy approach
in the lower orders it is convenient to use formulae \eq{feta}-\eq{fulldifhom}. The simplest way to generalize them is to replace
   the total differential  $\dr$ \eq{EBS+Q} by $\dr_Q $ \eq{tdgh2}.

  For instance, formula  \eq{feta}  acquires the form
   \be
\label{fetaQ}
f_\mu=  \eta\int_{  \tau\gs  } \mu(\tau,\gs )  \dr_Q \Go^\ga \dr_Q \Go_\ga\,
\Ex  G_l(g(r))\big |_{r=0}\,
\ee
with $\Go_\ga  $
\eq {Omega=}, $\Ex$ \eq{Exp} and
\be\label{fulldifhomQ}
\dr_Q =d z^\ga \ff{\p}{\p z^\ga} +d\tau\ff{\p}{\p \tau}
 + d\gs_i \ff{\p}{\p \gs_i}  \,
-\ff{1}{2i}  Q.\ee

Note that, although for our purpose $ \Go_\ga$ is taken in  the form \eq {Omega=}, generally,  it   can  also depend  on $c,b$
and $\gT$
while  the coefficients  $g$ must be in $\dr_Q-$cohomology. As a result,
the following counterpart of \eq{dGOGO=0} takes place \be\label{dGOGO=LQ2}\!\!
\dr_Q\big  ( \dr_Q \Go)^2
\Ex  \big)  =0\,.
\ee

\subsection{Examples}

To illustrate the   advantage of the proposed scheme
consider the evaluation of $S_1$  and $W_1$.

Equation \eq {EBS+Q} in this approach  acquires the form
\bee \label{EBS+QDH}  -2i \dr_Q S_1       \rule{0pt}{14pt} +i\eta\int_{\gt  } \dr\gt\gd(1-\gt) \theta_{\ga}\wedge \theta^{\ga}
   C *k *e^{i\gt z_{\ga}y^{\ga}}  +
i\bar\eta\int_{\bgt  } \dr\bgt\gd(1-\bgt)\bar\theta_{\pa}\wedge \bar\theta^{\pa}   C *\bar k*e^{i\bgt\bar z_{\pa}\bar
y^{\pa}} = 0 .\,\,\,\eee
Then, following \cite{Vasiliev:2023yzx},
one finds
  \bee\label{S1withwQ}
&&   S _1  =
 -\ff{\eta}{2 }\int_{\gt  } l(\gt)
  (\dr_Q\wmv )^2\EE(\wmv ) C * k-\ff{\bar{\eta}}{2 }\int_{\bgt  } l(\bgt)
(\dr_Q   \bwmv )^2
  \bEE(\bwmv) C * \bk\q
 \\\label{wS1withwQ}
&& \wmv^\ga= \gt z^\ga \q
 \bwmv^\pa=\bgt \bz^\pa \,.\eee
Since
\be\label{dQ}
\dr_Q   \wmv=
d\gt z^\ga+\gt\theta^\ga -\ff{i}{2} \gt c^{\gb\ga} z_\gb \,, \ee
one finds
 \be
\label{dQ2}(\dr_Q   \wmv )^2\cong 2 \gt d \gt z^\ga \theta_\ga
+i  \gt d\gt    c^{\ga \gb}z_\ga z_\gb  .\ee
(The   $d\gt$-independent   term of $(\dr_Q   \wmv )^2$    is weak.)
As a result, the substitution of  \eq{dQ2} into \eq{S1withwQ} yields
  \bee\label{S1withwQ=}
&&   S _1  =
 - {\eta} \int_{\gt  } l(\gt)
  \gt d \gt(  z^\ga \theta_\ga
+i \half c^{\gb \ga}z_\ga z_\gb) \EE(\wmv ) C * k+cc\,,
\eee
which coincides with $S_1$ \eq{S'+S''}.

Analogously,   Eq.\eq {W1goC} acquires the form  \bee
\label{W1goCQ}
\ls {W'}{} \,|_{\go C} &=& \ff{ i\eta}{4 }
\!\!\int\limits_{ \gt,   \gs  } l(\gs)    l(\gt)  (\dr_Q\wmv )^2
 \EE(\wmv)  \go  C  *k
+cc\eee
with $\wmv^\ga =\gt z^\ga-(1-\gt) (- \gs    p_\go{} )^\ga $.
Since \be(\dr_Q   \wmv )^2\cong 2 \gt(1-\gt)  d \gt  \dr \gs z^\ga  p_\go{}  _\ga\,    \ee
 ${W'|_{\go C}}$ \eq{W1goCQ} coincides with ${W'|_{\go C}}$ \eq {W1goC}.
For ${W'|_{C\go  }}$  one proceeds analogously.
  \section{Conclusion}
\label{Con}

In this paper,  an extension of the standard equations of the
HS gauge theory  is proposed  that makes local Lorentz symmetry of the model manifest. The main idea is to use its BRST extension {\it a la}
\cite{Vasiliev:2025hfh} with respect to the local {}$\mathfrak{sl}(2|\mathbb{C})$ Lorentz symmetry
in four dimensions. This leads to an additional Stueckelberg  symmetry trading
 the spin two Lorentz connection $\go^L$ of the original HS theory for the new
$\mathfrak{sl}(2|\mathbb{C})$ gauge field $\hat \go$, that allows one to impose the gauge condition
$\go^L=0$ yielding the  proper Lorentz covariant derivative $D^L$ with respect
to $\hat \go$ at all stages of the computations.
There are also new space-time two-form fields $U^{\ga\gb}$ and $\bar U^{\dga\dgb}$
associated with the expansion in powers of the BRST antighosts $b_{\ga\gb}$
and $\bar b_{\dga\dgb}$, that play a role of the Riemann tensor being
expressed in terms of the other fields by virtue of the gauge condition $\go^L=0$. The fields
$U^{\ga\gb}$ and $\bar U^{\dga\dgb}$ are analogous to the field tensors
$R^{\ga\gb}$ and $\bar R^{\dga\dgb}$ introduced in \cite{Didenko:2017qik}.

Note that the role of the BRST extension of the HS theory of this paper  differs from that of the usual
BRST formulations of HS theories both in the direct BRST description of the HS spectra
and vertices (see, {\it e.g.,} \cite{Bengtsson:1987jt}-\cite{Buchbinder:2024gll} and references therein) and in the approach
of \cite{Vasiliev:2025hfh}, where the BRST operator is constructed from a kind of Howe dual algebra to the HS algebra.
In the approach of this paper, the BRST
operator originates from an additional Lorentz algebra, that generates an additional Lorentz connection related to
the Lorentz component of the original HS field via a nonlinear
Stueckelberg symmetry. As such the BRST operator of this paper plays an auxiliary role and
cannot be used alone for the description of the theory as a whole.

The advantage of the
proposed scheme compared to that of \cite{Didenko:2017qik} is that it applies to any homotopy solution scheme of the HS equations
including, as  explained in the paper, the differential homotopy of \cite{Vasiliev:2023yzx}
 while the approach of \cite{Didenko:2017qik} was elaborated for the simplest conventional
(not shifted) homotopy.
The proposed general scheme is illustrated by the lower-order analysis of the theory to demonstrate its computational efficiency.
Let us stress that the proposed approach can be directly applied
to more general HS models in $AdS_3$ and $AdS_4$ including the extended system of \cite{Vasiliev:2015mka}, Coxeter HS theories of \cite{Vasiliev:2018zer}
(see also \cite{Tarusov:2025qfo}) and the models of \cite{different formalism5}.

\section*{Acknowledgements}

We would like to thank Sergey Fedoruk and Joseph Buchbinder for  useful discussions, Vyatcheslav Didenko and Nikita Misuna
for helpful comments on the manuscript
 and Per Sundell for the correspondence.
We are grateful for hospitality to Ofer Aharony,
Theoretical High Energy Physics Group of Weizmann Institute of Science where a part of this work was done.
This work was supported by Theoretical Physics and
Mathematics Advancement Foundation “BASIS” Grant No 24-1-1-9-5.

\end{document}